\documentclass[twocolumn,showpacs,preprintnumbers]{revtex4}
\usepackage{amsmath}

\input{tcilatex}

\begin{document}

\title{Energy landscape and rigidity}
\author{Gerardo G. Naumis}
\date{\today }
\pacs{64.70.Pf, 64.60.-i, 05.70.-a, 05.65.+b}

\begin{abstract}
The effects of floppy modes in the thermodynamical properties of a system
are studied. From thermodynamical arguments, we deduce that floppy modes are
not at zero frequency and thus a modified Debye model is used to take into
account this effect. The model predicts a deviation from the Debye law at
low temperatures. Then, the connection between the topography of the energy
landscape, the topology of the phase space and the rigidity of a glass is
explored. As a result, we relate the number of constraints and floppy modes
with the statistics of the landscape. We apply these ideas to a simple model
for which we provide an approximate expression for the number of energy
basins as a function of the rigidity. This allows to understand certains
features of the glass transition, like the jump in the specific heat or the
reversible window observed in chalcogenide glasses.
\end{abstract}

\maketitle

\address{$^{1}$Instituto de Fisica, Universidad Nacional Aut\'{o}noma de
M\'{e}xico (UNAM)\\
Apartado Postal 20-364, 01000, Distrito Federal, Mexico.}

\section{Introduction}

The physics of glass formation is a complex multiparticle problem, and in
spite of its importance from the fundamental and technological point of
view, many important questions remain unanswered \cite{Anderson}. As an
example we can cite the origin of the non-exponential relaxation laws \cite%
{Relaxation} or the ability of certain materials to reach the glassy state 
\cite{Jackle}. To tackle these problems there are many different approaches 
\cite{Debenedetti}: phenomenological models like the Gibbs-Dimarzio,
theoretical theories like the mode coupling or the use of extensive computer
simulations \cite{Debenebook}. A very interesting question is how the glass
transition temperature ($T_{g}$) depends on chemical composition.
Chalcogenide glasses (formed with elements from the VI column doped with
impurities) are very useful for understanding these effects \cite{Boolchand1}%
. As was discovered more than 2,000 years ago, $T_{g}$ can be raised or
lowered by adding impurities, and the fragility of the glass can be changed
from strong to fragile \cite{Tatsumisago}. Recently, by using stochastic
matrices \cite{Kerner1,Kerner2}, the law that gives the relation between $%
T_{g}$ and the concentration of modifiers \cite{Sreeram} has been obtained,
including a constant that appears in the law for almost any chalcogenide
glass \cite{Micoulaut}. Another interesting property of glasses is the
behavior of their viscosity, which is usually referred as fragility \cite%
{Tatsumisago}.The fragility of a glass is also related with the ease of
glass formation, in the sense that strong glasses are those that do not
require a high speed of cooling, and fragile glasses are weak glass formers
that require a rapid quench. The ease of glass formation can be explained at
least in a qualitative way by the rigidity theory (RT), introduced by
Phillips \cite{Phillips1} and further refined by Thorpe \cite{Thorpe0}. By
considering the covalent bonding as a mechanical constraint, the ease of
glass formation is related with the ratio between available degrees of
freedom and the number of constraints. If the number of constraints is lower
than the degrees of freedom, there are zero frequency vibrational modes
called floppy \cite{Thorpe1}. The resulting network is under-constrained. A
transition occurs when the lattice becomes rigid. Glasses that are rigid due
to a certain chemical composition are easier to form, and many features of
this transition have been experimentally observed \cite{Boolchand1}\cite%
{Boolchand2}. Even for simple systems like hard-disks \cite{Huerta0} and
colloids \cite{Huerta1}, it seems that rigidity plays an important role .
For more complex systems like proteins, rigidity has been used as a very
powerful tool to understand folding and long-time scale motions \cite%
{Brandon}.

A very puzzling fact of RT that has not been explored is the following:
according to the idea of looking at rigidity as a vectorial percolation
problem, at the rigidity threshold the entropy is high \cite{Moukarzel}, due
to strong fluctuations as happens in any phase transition. One even can
define a free energy and specific heat as a function of the flexibility of
the system, that has a singularity at the transition \cite{Duxbury}.
However, the experimental data from modulated scanning calorimetry in
chalcogenide glasses shows the opposite: at the rigidity transition the
configurational entropy is less and there is a \textit{window of
reversibility }\cite{Boolchand1}\cite{Georgiev}. Specially, it has been
observed that protein folding is reversible because it occurs at the
rigidity transition \cite{Brandon}, and this seems to be a crucial property
for life to exist \cite{Brandon}.

Mainly, the problem resides in the fact that although RT provides a
framework to understand many features of a system, its use in a quantitative
way has not been fully developed to provide a link with the thermodynamics
of the system \cite{Jacobs}. In a previous paper we approached this problem
by using a phenomenological free energy to account for many thermodynamical
properties of the glass transition \cite{Naumis}, and then we made extensive
computer simulations with associative fluids to show that many concepts of
the RT work in a "thermodynamical environment" \cite{Huerta}\cite{Huertaprb}%
. However, the connection with thermodynamics is still not mature, since
there is no general way of introducing thermodynamics in the RT.

In a different context, the energy landscape is a formalism that has been
very useful for describing the molecular scale events that happen during the
glass transition \cite{Angell}. The landscape is a multidimensional surface
generated by the system potential energy as a function of the molecular
coordinates \cite{Debenedetti}. In an $N$ body system the landscape is thus
determined by the potential energy function, given by $\Phi (\mathbf{r}%
_{1},...,\mathbf{r}_{N})$ where $\mathbf{r}_{i}$ comprise position,
orientation and vibration coordinates. For the simplest case of particle
possessing no internal degrees of freedom, the landscape is a $(3N+1)$
object. The topography of this landscape is fundamental for the
thermodynamics of the system. At high temperatures the system does not feel
the summits and valleys of $\Phi (\mathbf{r}_{1},...,\mathbf{r}_{N})$
because the kinetic energy contribution dominates. However, as the
temperature is lowered the system is unable to surmount the highest energy
barriers and therefore is forced to sample deep minima. When this happen,
the kinetic of relaxation changes from exponential to stretched exponential 
\cite{Debenebook}. An important observation is that, according to
statistical mechanics, the entropy of the system depends on the accessible
volume in the phase space. However, inside a local minimum of the potential
energy, it can happen that if there are no paths that connect to other
minima, the system cannot sample that part of the phase space. In such a
case, ergodicity is broken and the system is no longer in thermal
equilibrium. Such a glass will have a residual entropy \cite{Goldstein}. In
this article, we show that rigidity can be related with the statistics of
the energy landscape, since the number of floppy modes is related to the
number of different configurations of the system with nearly equal minimal
energies, and thus provides an estimation for the number of minima energy
basins of the landscape. But floppy modes also provides channels in phase
space that increase the entropy, which in part explains the paradox of the
window of reversiblity. To show these connections, we will concentrate in
the effects of rigidity on the shape of the energy landscape.

The layout of this work is the following: in section II we discuss a simple
way to introduce thermodynamics into RT, however, as we will see, the
straightforward manner of doing this do not agree with the experimental
results. Thus we propose that the effects of floppy modes are only important
at low temperatures or during glass transition. In section III the
connection with the energy landscape is made and a simple model is worked
out. Finally, in section IV we give the conclusions.

\section{ Rigidity and thermodynamics}

In this section we explore some simple thermodynamical consequences of the
RT. As explained before, the rigidity ideas of Phillips \cite{Phillips2} and
Thorpe \cite{Thorpe1} were used in order to understand the ease of glass
formation. In this theory, the ability for making a glass is optimized when
the number of freedom degrees, in this case $3N,$ where $N$ is the number of
particles, is equal to the number of mechanical constraints ($N_{c}$) that
are given by the bond length and angles between bonds.

The number $(3N-N_{c})/3N$ gives the fraction of cyclic variables of the
Hamiltonian, \textit{i.e.}, when one of these variables is changed, the
energy of the system does not change, as for example happens with the center
of mass coordinate. This fraction also corresponds to the fraction of
vibrational modes with zero frequency ($f$), called floppy modes,\ with
respect to the total number of vibrational modes. The counting of floppy
modes in a mean-field, known as Maxwell counting, goes as follows \cite%
{Thorpe2}: since each of the $r$ bonds in a site of coordination $r$ is
shared by two sites, there are $r/2$ constraints due to distance fixing
between neighbors. If the angles are also rigid, in $3D$ there are $(2r-3)$
constraints, to give, 
\begin{equation*}
f=\frac{3N-N_{c}}{3N}=1-\sum_{r}\frac{\left[ r/2+(2r-3)\right] x_{r}}{3}=2-%
\frac{5}{6}\left\langle r\right\rangle
\end{equation*}%
where the last term corresponds to the angular constraints, $x_{r}$ is the
fraction of particles with coordination $r$, and $\left\langle
r\right\rangle $ is the average coordination number, defined as, 
\begin{equation*}
\left\langle r\right\rangle =\sum_{r}rx_{r}
\end{equation*}%
A rigidity transition occurs when $f=0$ and the system pass from a floppy
network to rigid one. If $f$ is a negative number, \textit{i.e.}, if there
are more constraints than degrees of freedom, the lattice is overconstrained
and the important number is how many stressed bonds are present. In $3D$,
the rigidity transition leads to the critical value $\left\langle
r_{c}\right\rangle =2.4$ if all angular constraints are considered. In real
systems, the Maxwell counting breaks near the rigidity transition, and the
number of floppy modes are obtained form the pebble game algorithm \cite%
{Thorpe1}.

What are the simple thermodynamical effects of floppy modes? To answer this
question, first we use the most simple model for atomic vibrations in the
harmonic approximation, where the interatomic potentials are replaced by
springs. The corresponding Hamiltonian is,%
\begin{equation}
H=\sum_{j=1}^{3N}\frac{P_{j}^{2}}{2m}+\sum_{j=1}^{3N(1-f)}\frac{1}{2}m\omega
_{j}^{2}Q_{j}^{2}  \label{hamiltonian}
\end{equation}%
where $Q_{j}$ and $P_{j}$ is the $j-$normal mode coordinate in phase space,
and $\omega _{j}$ is the corresponding eigenfrequency of each normal mode.
Observe that floppy modes have zero frequency; they do not contribute to the
elastic energy. Using simple statistical mechanics, we can obtain the
partition function in the canonical ensemble at the classical limit (high
temperatures compared with the Debye temperature), 
\begin{eqnarray*}
Z &=&\didotsint \dprod\limits_{j=1}^{N}dP_{j}dQ_{j}e^{-H/kT} \\
&=&\left( \frac{2\pi mkT}{h^{2}}\right) ^{\frac{3N}{2}}\dprod%
\limits_{j=1}^{3N(1-f)}\left( \frac{2\pi kT}{m\omega _{j}^{2}}\right) ^{%
\frac{1}{2}},
\end{eqnarray*}%
where $T$ is the temperature and $k$ the Boltzmann constant. The free energy
of the system is now given by,%
\begin{equation*}
F=-\frac{3NkT}{2}\ln \left( \frac{2\pi mkT}{h^{2}}\right) -\frac{kT}{2}%
\sum_{j=1}^{3N(1-f)}\ln \left( \frac{2\pi kT}{m\omega _{j}^{2}}\right) .
\end{equation*}%
From this last expression, the corresponding specific heat ($C_{V}$) is,%
\begin{equation*}
C_{V}=3Nk-\frac{3Nk}{2}f.
\end{equation*}%
In this simple approach, the prediction is that $C_{V}$ is given by the
Dulong-Petit law, minus a term that depends on the number of floppy modes.
The reason is clear: \textit{floppy modes do not store energy} since they
are cyclic variables of the Hamiltonian, as for example happens with the
center of mass. However, a careful examination of the experimental data
shows that for chalcogenide glasses \cite{Boolchand1}, like compounds of $%
As-Ge-Se$, and ferroelectric materials \cite{Singh}, $C_{V}$ does not
depends on $f$. Instead, they follow the Dulong-Petit law. From this simple
thermodynamical argument, one is lead to propose that floppy modes do not
have a perfect zero frequency,\textit{i.e.}, in real glasses they are
shifted by residual forces, like the Van der Waals interaction. This
argument is confirmed by neutron scattering experiments, where it has been
shown that floppy modes in $As-Ge-Se$ are blue-shifted \cite{Boolchand2},
forming a peak at around $5meV.$ Thus, at high temperatures, all the $3N$
oscillators are excited. Only at low temperatures we suggest that the
effects of floppy modes are important since all floppy modes are frozen
nearly at the same temperature. The corresponding temperature ($\Theta _{f}$%
) where these modes are frozen, can be estimated from the energy required to
excite modes of $5meV$, that gives $\Theta _{f}\sim 60%
%TCIMACRO{\U{b0}}%
%BeginExpansion
{{}^\circ}%
%EndExpansion
K.$

This behavior at low temperatures, where a quantum treatment is needed, can
be modelled by using a simple density of states $\rho (\omega )$ that takes
into account the floppy peak in the spectrum. First we use a Debye type of
density of states, normalized to $3N(1-f).$Then we add the contribution from
the floppy modes, with a delta function centered around a characteristic
peak at $\omega _{0}$. The corresponding density of states is,%
\begin{equation*}
\rho (\omega )=\left\{ 
\begin{array}{c}
\frac{9N(1-f)}{\omega _{D}^{3}}\omega ^{2}+3Nf\delta (\omega -\omega _{0}),%
\text{ if }\omega \leq \omega _{D} \\ 
0\text{ if }\omega >\omega _{D}%
\end{array}%
\right.
\end{equation*}%
where $\omega _{D}$ is the Debye cut-off frequency. By using the
Bose-Einstein distribution for the number of phonons in equilibrium at a
certain temperature, we get that the specific heat is,%
\begin{equation*}
C_{V}=(1-f)3NkD(x_{0})+f3Nk\frac{x^{2}e^{x}}{(e^{x}-1)^{2}}
\end{equation*}%
where $x=\Theta _{fl}/T,$ $x_{0}=\Theta _{D}/T$, and $\Theta _{D}=\hbar
\omega _{D}$ is the Debye temperature. $D(x_{0})$ is the well known Debye
function. At high temperatures, the model predicts the Dulong-Petit law as
expected, while at low $T$, the following behavior is obtained,%
\begin{equation*}
C_{V}\approx (1-f)3Nk\frac{4\pi ^{4}}{5}\left( \frac{T}{\Theta _{D}}\right)
^{3}+f3Nk\left( \frac{\Theta _{f}}{T}\right) ^{2}e^{-\frac{\Theta _{f}}{T}}.
\end{equation*}%
which is a Debye law of the type $T^{3},$ but with a contribution that is in
the form of the Einstein model. Each contribution is determined from the
fraction of floppy modes for a given composition of the glass. The present
model suggest that experiments at low temperatures performed on chalcogenide
glasses will provide caracteristic features of rigidity.

\bigskip

\section{Energy landscape and rigidity}

\bigskip

In the last section we discussed that floppy modes have effects mainly at
low $T$. In spite of this, an examination of the experimental results shows
that the number of floppy modes is also important for the thermodynamical
properties at the glass transition \cite{Tatsumisago}. For example, the
magnitude in the jump of $C_{P},$ usually denoted by $\Delta C_{P}$, the
jump in the thermal expansion, the energy for activation of viscosity, the
fragility and the entropy of a liquid melt depends on $f$. Moreover, very
recently the group of Boolchand discovered the window of reversibility in
the heat flow, associated with a phase of zero internal stress in the
lattice \cite{Georgiev}. Angell has pointed out the qualitative relationship
between energy landscape and fragility during glass transition \cite{Angell}%
. However, still is no clear how to relate these features with the
statistics of the landscape. Here we will show that the number of floppy
modes provides a useful parameter to represent the roughness of the
landscape. This roughness is evident when the glass is melted explaining why
floppy modes are important during glass transition, since they are
collective motions that provide pathways across the phase-space and energy
landscape.

As a first and tentative step, we start again by supposing that floppy modes
are at zero frequency. Around any given inherent structure, the potential
has a minimum and thus can be expanded in a Taylor series, which turns out
to be the expression of an harmonic potential. From the Hamiltonian
presented in eq.(\ref{hamiltonian}) is clear that in a inherent structure,
each floppy mode provides a \textit{channel in the landscape} since the
energy does not depend upon a change in a floppy coordinate. A very simple
example is shown in fig. 1, which shows the bottom of the landscape for a
system with two normal modes. In the first system, (Fig. 1a) $f=0,$ but the
other has $f=1$ (Fig. 1b) since one of the springs constants was set to zero
(of course, by excluding the center of mass coordinate). In a more general
way, for a given inherent structure, the number of channels is clearly given
by $f.$ Each channel increases the available phase space allowed to visit.
The entropy due to floppy modes is easy to calculate. In the microcanonical
ensemble, the number of accessible states ($\Omega (E,V,N)$) for a system
with a volume $V$ is proportional to the area defined by the surface of
constant energy, $E=H(P_{1},...P_{N},Q_{1},...,Q_{N}).$ Since floppy modes
are cyclic variables of the Hamiltonian, we can write,%
\begin{equation*}
\Omega (E,V,N)=\underset{E=H(P_{1},...,Q_{3N(1-f)})}{\frac{1}{h^{3N}}\int
...\int }\dprod\limits_{j=1}^{3N}dP_{j}\dprod\limits_{k=1}^{3N(1-f)}dQ_{k}%
\left( \int_{0}^{V^{1/3}}dQ\right) ^{3Nf},
\end{equation*}%
and using the Boltzmann relation $S=k\ln \Omega (E,V,N)$ we get,%
\begin{equation}
S=\ln \left[ \frac{\left( 2\pi m\right) ^{\frac{3N}{2}}E^{3N(1-f/2)}}{h^{3N}%
\left[ (3N(1-f/2))-1\right] !}\dprod\limits_{j=1}^{3N(1-f)}\left( \frac{2}{%
m\omega _{j}^{2}}\right) \right] +fNk\ln V  \label{entropy}
\end{equation}%
The entropy provided by the channels in the landscape is simply given by the
last term, $S_{c}=fNk\ln V.$ At first glance, it seems that this result
agrees with the experimental observations, because during glass transition,
it has been observed that floppy glasses have a great entropy and as a
result, they have a more fragile behavior as deduced from the Adams-Gibbs
relation \cite{Tatsumisago}. However, a more detailed analysis shows that if
we suppose an entropy of the type given by eq. (\ref{entropy}), the specific
heat does not follow the Dulong-Petit law. This is due to the dependence of $%
S$ upon $E^{3N(1-f/2)}$, which is just a result of the independence of $H$
with respect to floppy modes. As discussed in the previous section, this
leads to the conclusion that floppy modes are not strictly at zero
frequency. The blue-shift of the floppy modes means that the \textit{%
channels in phase space are not flat}: there is a small curvature in the
direction of the floppy variable. This effect has the property that it
restores the Dulong-Petit law and provides directions in phase space where
the system can relax without big changes in energy.

In a floppy glass there is a hierarchy in the strength of the forces. The
forces that restores the Dulong-Petit law are the weakest. Then it is
natural to assume that the anharmonic contributions of these residual forces
are also small. Under this assumption, the extra entropy due to these modes
is $S\simeq fNk\ln V$ which is only activated when the glass traverses the
glass transition. Furthermore, we can speculate that these channels are in
fact the ones that explains the fragility and ease of glass formation since
is clear that is much more difficult to trap the system in a local minima of
the landscape when many channels are present.

However, there are two important facts to consider in all the previous
statements: first the number of floppy modes is a function of the energy. In
fact, when the glass becomes fluid, most of the constrictions upon the bond
lengths and angles are relaxed and $f$ is raised. For the extreme case of
no-bonding between atoms, the system behaves without constraints and all the
modes are floppy $f=1$. Notice that an ideal gas is a perfect "floppy
system". An improvement to eq.(\ref{entropy}) is to make $f$ a function of $%
E,$ then the number of floppy modes is $3Nf(E).$In such a case, the jump in
the specific heat will also depend on $f$, as observed in the experiments.
The function $f(E)$ is zero when $E\gg kT_{g}$ and has a value determined by
the average coordination number below the glass transition,i.e., $f(E)=2-%
\frac{5}{6}\left\langle r\right\rangle .$ The shape of this function can be
estimated using a procedure that we will describe later.

The second consideration is that the number of floppy modes affects the
number of minima energy valleys (usually called inherent structures) that
are available when the system has a certain energy. This effect is explained
in figure 2, where a system of bars and hinges is considered. In the example
of figure 2,\ there are no angular forces. Each bar provides a restriction
to the system. There are three squares. In one of the squares there is a
diagonal bar. As a result, this square can not be deformed, since the
distance between all the hinges are fixed. The other two squares are
flexible as indicated by the arrows. Each of these flexible squares can be
deformed independently, and the system has $2$ floppy modes (again, without
counting the center of mass translation and rotations around it). Now we
move the diagonal bar to the second square and the system has the same
number of floppy modes, but the structure is different, and the same thing
happen if we put the diagonal in the first square. In the landscape
formalism, each of these configurations is in a different "inherent
structure" and corresponds to a basin with the same energy. This part of the
entropy has been studied extensively in the context of rigidity transitions 
\cite{Moukarzel}. However, as we will see next, there is a competition
between the channel and configurational entropies.

To see how these concepts are applied in a particular case, let us consider
the following two dimensional model that contains all of the previous
features that we discussed. Consider a system of $N$ disks interacting with
a central force where no angular forces are considered. Each disk has a hard
core potential and an attractive part which has a range determined by the
parameter $\lambda .$ If $\sigma $ is the diameter of the disks, $r$ is the
distance between the centers of two disks, the potential is written as,%
\begin{equation*}
V(r)=\left\{ 
\begin{array}{c}
\infty \text{ \ \ if \ \ }r<\sigma \\ 
-V_{1}\text{ \ \ \ \ \ if \ }\sigma \leq r\leq \lambda \sigma \\ 
0\text{ \ \ if \ }r>\lambda \sigma%
\end{array}%
\right.
\end{equation*}

The nature of the fluid and solid phase of this system has been studied in a
previous work \cite{Huerta1}. Here we only study the rigidity. Within this
model, a bond is formed when the distance between two disks is between $%
\sigma $ and $\lambda \sigma .$ Each bond has an energy $-V_{1}$, and the
energy of the system is just proportional to the number of bonds. This
number is proportional to the average coordination number divided by $2$
since each bond is shared by two sites. Then, the amount of energy ($E$) of
the system is given by,%
\begin{equation}
E=-V_{1}N\frac{\left\langle r\right\rangle }{2}\simeq -2V_{1}N(1-f)
\label{E(f)}
\end{equation}%
where it was used that for the mean field approximation in two dimensions $%
f\simeq (2N-(N<r>/2))/2N$. From the last equation, is clear that a gas is
obtained when the system is $100\%$ flexible ($f=1)$ and the state of
maximal packing (the hexagonal lattice with maximal coordination $r_{\max }=6
$) is overconstrained (there are $N/2$ redundant bonds in the mean field
approximation). Notice that $f$ is a function of $E.$

As said previously, there is an entropy provided by floppy modes channels ($%
S_{1}$) and to the different configurations of floppy modes ($S_{2}$).
According with our previous assumptions, the first contribution is $%
S_{1}\simeq fNk\ln A$ where $A$ is the area of the system. This is only
valid in the flexible phase,\textit{\ i.e}., before the freezing of the
system since at that point it has been suggested that there is a rigidity
transition \cite{Huerta}. After freezing, this contribution is zero ($%
S_{1}=0).$ At high temperatures, the system is a fluid and the entropy is
just the same as the one obtained from the available phase space without any
interaction. A more realistic assumption although still very rough is to use
that $S_{1}\simeq fNk\ln (A-b)$ where $b$ is proportional to the area
occupied by the disks \cite{Tabor} $b\approx N\pi (\lambda \sigma )^{2}/2$.

The other contribution to the entropy comes from the number of ways in which
a configuration with a given $\left\langle r\right\rangle $ can be made.
Although this number is difficult to calculate, one can suppose a cell model
of the fluid, and then just consider the number of ways in which absent
bonds can be deleted from the lattice with maximal packing$.$ This number of
configurations ($\Omega (f,N)$) is,%
\begin{equation}
\Omega (f,N)=\frac{\left( \frac{r_{\max }N}{2}\right) !}{\left( \frac{%
r_{\max }N}{2}-\frac{\left\langle r\right\rangle N}{2}\right) !\left( \frac{%
\left\langle r\right\rangle N}{2}\right) !}  \label{omega}
\end{equation}%
where $\left\langle r\right\rangle $ is a function of $f.$ The corresponding
configurational entropy is $S_{2}=k\ln \Omega (f,N)$. A natural way to
compute this entropy is to define an order parameter $m(f)$ as,%
\begin{equation}
m(f)=\frac{(r_{\max }-\left\langle r\right\rangle )-\left\langle
r\right\rangle }{r_{\max }}\simeq \frac{4f-1}{3}  \label{m}
\end{equation}%
In terms of this parameter, and using Stirling%
%TCIMACRO{\U{b4}}%
%BeginExpansion
\'{}%
%EndExpansion
s approximation, the total entropy for $f\geq 0$ now reads,%
\begin{eqnarray*}
\frac{S_{1}+S_{2}}{Nk} &=&\ln 2+f\ln (A-b)- \\
&&\frac{(m(f)+1)}{2}\ln (1+m(f))-\frac{(1-m(f))}{2}\ln (1-m(f))
\end{eqnarray*}%
For an overconstrained lattice, the expression for the entropy is just $S_{2}
$ \ The expression for $f\geq 0$ contains the effects that were discussed
previously, \textit{i.e.}, the linear dependence of the entropy upon $f$,
and the contribution from different structures with the same energy. In
figure 3 we show a plot of the total entropy and the corresponding
contributions for a given $A-b$. It is interesting to note that $S_{2}$
tends to grow as we diminish the number of floppy modes, since the number of
configurations with the same energy grows. Notice that $S_{2}$ does not have
a maximum exactly when $f=0$ due to the mean field approximations; the
maximum is shifted to the right. From eq. (\ref{m}), this occurs near $%
\left\langle r\right\rangle =4,$.i.e., near the two dimensional rigidity
transition. This fact seems to contradict that in the rigidity transition,
the experimental non-reversible heat flow is a minimum, which means that the
configurational entropy is a minimum. One can expect that in the rigidity
transition, a lot of fluctuations will be observed, while in the experiments
it seems that the contrary is true \cite{Tatsumisago}. However, the present
results shows that floppy modes have \textit{two competing effects,} one is
the entropy due to the different configurations, but the other is the shape
of each basin, since around inherent structure, floppy modes form channels
that increase the entropy. Thus, as is shown in figure 3, when the system
pass from flexible to rigid, the number of configurations raises, but the
number of channels diminish . Experimental results suggests that this last
effect is more important, since the configurational entropy of a melt with a
floppy glass former is higher as the number of floppy modes is increased 
\cite{Tatsumisago}\cite{Boolchand2}.

Finally, the free energy for $f\geq 0$ can be written as,%
\begin{equation}
\frac{F(f)}{Nk}=-2V_{1}(1-f)\frac{V_{1}}{k}-T(S_{1}+S_{2})  \label{F(f)}
\end{equation}

To compare with the energy landscape formalism, we use that the partition
function is the sum of partition functions at inherent structures\cite%
{Buchner},%
\begin{equation}
Z(T)=Z^{ha}(T)\int_{0}^{\infty }G(E)e^{-E/kT}dE  \label{zaux}
\end{equation}%
where $G(E)$ is the number of energy basins with energy $E$, and $%
Z_{i}^{ha}(T)$ is the partition function for a system of harmonic
oscillators \cite{Buchner}. Since $G(E)$ is always a growing function, and $%
e^{-E/kT}$ is always decreasing, the integral of eq.(\ref{zaux}) can be
replaced by the value at the maximum $\overline{E}$,%
\begin{equation*}
\int_{0}^{\infty }G(E)e^{-E/kT}dE\approx G(\overline{E})e^{-\overline{E}/kT}
\end{equation*}%
The corresponding free energy $F$ is,%
\begin{equation}
\frac{F}{NkT}=-\ln Z(T)=-\ln Z^{ha}(T)-\ln G(\overline{E})+\frac{\overline{E}%
}{kT}  \label{F}
\end{equation}%
As usual, the free energy is just the contribution from the vibrations
inside the basin, the entropic component due to the existence of different
basins, and an energetic component which reflects the average depth of the
landscape at a certain $T$. Now we turn our attention in how $G(\overline{E}%
) $ is affected by the floppy modes. Comparing eq.(\ref{F}) and eq.(\ref%
{F(f)}), we get an estimation for $G(\overline{E}),$ 
\begin{equation}
G(\overline{E})\approx \exp N\left[ -m(\overline{E})\ln (A-b)-\frac{(m(%
\overline{E})+1)}{2}\ln (1+m(\overline{E}))-\frac{(1-m(\overline{E}))}{2}\ln
(1-m(\overline{E}))\right]  \label{g(E)}
\end{equation}%
where $m(\overline{E})$ is obtained from eq.(\ref{E(f)}) and eq.(\ref{m}),%
\begin{equation*}
m(\overline{E})=1-2f(\overline{E})=\left( 1+\frac{2\overline{E}}{3V_{1}N}%
\right)
\end{equation*}%
If the channel term is the most important, $G(\overline{E})$ can be
approximated by,%
\begin{equation*}
G(\overline{E})\approx \exp \left[ N\left\vert m(\overline{E})\right\vert
\left( \ln A-(b/A)\right) \right]
\end{equation*}%
which has the same general shape of that proposed by Stillinger \cite%
{Stillinger}. The factor $\left\vert m(\overline{E})\right\vert \left( \ln
A-(b/A)\right) $ in the exponential can be identified with the landscape
complexity \cite{Fyodorov}.

\section{Conclusions}

In this article, we have explored the effects of floppy modes into the
thermodynamics of glasses. In particular, we shown that a blue-shift of
floppy modes can be predicted using simple thermodynamical arguments. This
leads to the formulation of a simple model, which suggests effects of floppy
modes at low temperatures. During glass transition, floppy modes also play a
role. Thus we explored how flexibility and rigidity determine the energy
landscape. We found two competing effects that contribute to the entropy in
the liquid melt; one contribution is given by channels and the other is the
existence of different energy basins. By considering a simple example, we
showed how to estimate both contributions, and we discussed the  effects in
the window of reversibility. The results of this article seem to confirm the
Phillip%
%TCIMACRO{\U{b4}}%
%BeginExpansion
\'{}%
%EndExpansion
s idea that glass forming tendency is enhanced at the rigidity transition 
\cite{Phillips3}, since although there is an increase in the entropy due to
the different energy basins, the pathways provided by floppy modes are
absent and the system is easier to trap in a certain minimum.\ 

\textbf{Acknowledgments.} This work was supported by DGAPA UNAM project
IN108502, and CONACyT-NSF joint project 41538.

\textbf{Figure 1}. Bottom of the landscape for a system a) with no floppy
modes and a potential energy given by $\phi (x,y)=x^{2}+y^{2}$ \ b) with one
floppy mode obtained by removing a "spring", $\phi (x,y)=x^{2}.$A channel is
generated in the $y$ direction.\FRAME{ftbpF}{2.2693in}{4.2999in}{0pt}{}{}{%
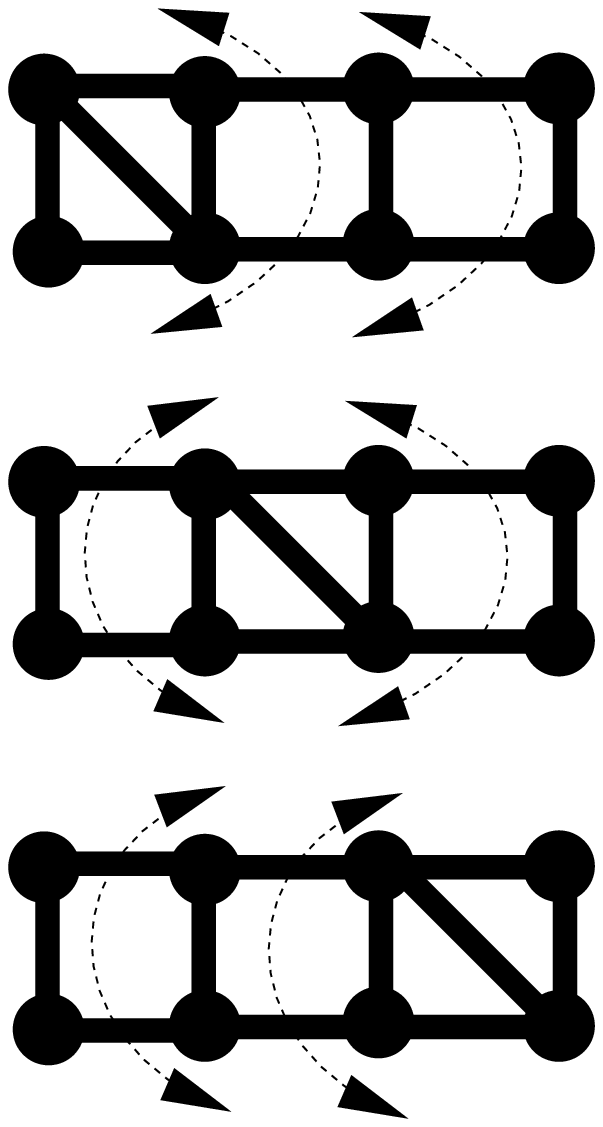}{\special{language "Scientific Word";type
"GRAPHIC";maintain-aspect-ratio TRUE;display "USEDEF";valid_file "F";width
2.2693in;height 4.2999in;depth 0pt;original-width 2.3315in;original-height
4.4521in;cropleft "0";croptop "1";cropright "1";cropbottom "0";filename
'fig2.eps';file-properties "XNPEU";}}

\textbf{Figure 2.} A system of bars and hinges with three different
configurations. The squares with the diagonal bars are rigid, while the
others are flexible. The corresponding floppy modes are shown with arrows.

\textbf{Figure 3.} Contributions to the total entropy (crosses) in units of $%
Nk$. The dotted line is the contribution from\FRAME{ftbpF}{2.3583in}{3.2534in%
}{0pt}{}{}{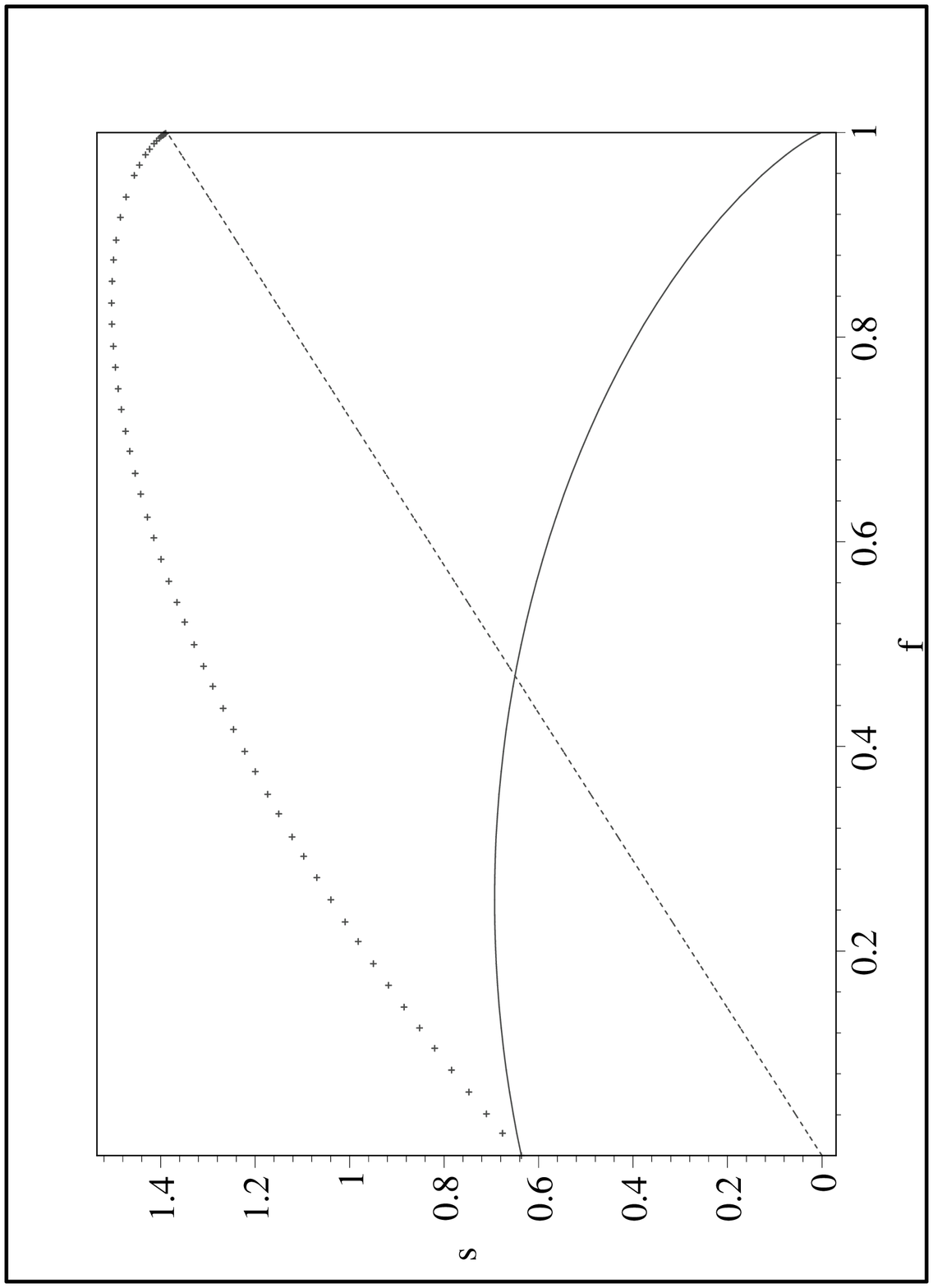}{\special{language "Scientific Word";type
"GRAPHIC";maintain-aspect-ratio TRUE;display "USEDEF";valid_file "F";width
2.3583in;height 3.2534in;depth 0pt;original-width 6.4913in;original-height
8.9854in;cropleft "0";croptop "1";cropright "1";cropbottom "0";filename
'fig3.eps';file-properties "XNPEU";}} channels ($S_{1}$) with the aribitrary
value $A-b=4$. The solid line is the contribution from different
configurations ($S_{2}$).

\end{document}